# TRB PAPER # 17-00238

## Can Data Generated by Connected Vehicles Enhance Safety?
## A proactive approach to intersection safety management

Mohsen Kamrani
Graduate Research Assistant, Department of Civil & Environmental Engineering
The University of Tennessee
mkamrani@vols.utk.edu

Behram Wali
Graduate Research Assistant, Department of Civil & Environmental Engineering
The University of Tennessee
bwali@vols.utk.edu

Asad J. Khattak, Ph.D.
Beaman Distinguished Professor, Department of Civil & Environmental Engineering
The University of Tennessee
akhattak@utk.edu

March, 2017

**Can Data Generated by Connected Vehicles Enhance Safety?**
**A proactive approach to intersection safety management**

Mohsen Kamrani, Behram Wali, Asad J. Khattak
The University of Tennessee, Knoxville

**Abstract –** Traditionally, evaluation of intersection safety has been largely reactive, based on historical crash frequency data. However, the emerging data from Connected and Automated Vehicles (CAVs) can complement historical data and help in proactively identify intersections which have high levels of variability in instantaneous driving behaviors prior to the occurrence of crashes. Based on data from Safety Pilot Model Deployment in Ann Arbor, Michigan, this study developed a unique database that integrates intersection crash and inventory data with more than 65 million real-world Basic Safety Messages logged by 3,000 connected vehicles, providing a more complete picture of operations and safety performance of intersections. As a proactive safety measure and a leading indicator of safety, this study introduces location-based volatility (LBV), which quantifies variability in instantaneous driving decisions at intersections. LBV represents the driving performance of connected vehicle drivers traveling through a specific intersection. As such, by using coefficient of variation as a standardized measure of relative dispersion, LBVs are calculated for 116 intersections in Ann Arbor. To quantify relationships between intersection-specific volatilities and crash frequencies, rigorous fixed- and random-parameter Poisson regression models are estimated. While controlling for exposure related factors, the results provide evidence of statistically significant (5% level) positive association between intersection-specific volatility and crash frequencies for signalized intersections. The implications of the findings for proactive intersection safety management are discussed in the paper.

Keywords: Proactive Safety, Driving Volatility, Crash Waiting to Happen, Connected Vehicles, Basic Safety Messages, Big Data, Fixed and Random Parameters, Poisson Regression

# INTRODUCTION

There is considerable evidence about vehicle conflicts at intersections resulting in crashes, making them among the most dangerous locations on roadways (1; 2). Traditionally, intersection safety evaluations are done based on historical data and they are largely reactive i.e. the state-of-the-art methods characterize unsafe intersections based on historical and expected crash frequencies (2; 3). Safety treatments can then be applied to intersections based on historical crash data methodology. Variability in instantaneous driving behaviors can be leading indicators of occurrence of unsafe outcomes such as crashes/incidents. In this study, we posit that expanding the concept of driving volatility (4-6) to specific locations (termed as Location-Based Volatility) by using real-world large-scale connected vehicle data has a significant potential in unveiling critical relationships between extreme driving behaviors (and its fluctuations) and safety outcomes at specific intersections.

The Safety Pilot Model Deployment (SPMD) offers detailed and relevant data. This pilot is underway in Ann Arbor, Michigan, intended to demonstrate vehicle-to-vehicle (V2V) and vehicle-to-infrastructure (V2I) communication in a real-world environment. Within SPMD, Basic Safety Messages (BSMs) contain rich information packets (exchanged at the frequency of 10 Hz) that describe a vehicle's position, motion, its component status, and other relevant information exchanged between vehicles/infrastructure through V2V and V2I applications (*7*). Such emerging data has been used for creating trip-based driving volatilities for drivers, capable of identifying abnormal or extreme behaviors prior to unsafe outcomes such as crashes/incidents (*6*). Important in this aspect is the concept of "driving volatility" that captures the extent of variations in driving, especially hard accelerations/braking, jerky maneuvers, and frequent switching between different driving regimes (*4*). Specifically, Wang et al. (*5*) and Liu and Khattak (*6*) examined the relationships between trip-based driving volatility and several factors such as demographics, trip purpose, duration, and detailed vehicle characteristics (*5; 6*). The potential of driver-specific trip-based volatilities for developing advanced traveler information systems, driving feedback devices, and alternative fuel vehicle purchase decision tools were concluded (*5; 6*).

This study focuses on developing an analytic methodology to examine instantaneous driving behaviors at specific locations, and its variability. The paper explores how variability in driving can be mapped to historical safety outcomes such as crashes at specific locations. Such an analysis is fundamental towards proactive intersection safety management.

## LITERATURE REVIEW

There are different branches of ongoing research topics in the connected vehicles (CV) area. Several major directions of research can be identified. Topics such as network robustness and information propagation efficiency (*8*) are still under investigation in order to establish a better vehicle to vehicle (V2V) and vehicle to infrastructure (V2I) connection (*8*). Another is the systems and algorithms whose ultimate goal are the reduction of the gap between vehicles in order to increase roads capacity and reduction in fuel consumption through different method such as speed harmonization (*9*), trajectory optimization (*10*) and platooning as discussed in Bergenhem et al. (*11*).

Also, there are a number of studies (not necessarily in CV area) trying to characterize aggressive, reckless or risky driving style (*12*). Among them, speed limits are usually the threshold that determines a driver's performance (*13; 14*). While characterizing driver's performance, the important finding is that risky driving behaviors have been found to be positively correlated with the likelihood of crashes or near-crash events (*15*). This said, a broad spectrum of studies related to connected vehicle systems have proposed mechanisms for warnings or alerts to drivers using the CV applications and their effect on safety. For instance, Chrysler et al. (*16*) investigated the effect of warning messages on drivers' ability to handle primary and secondary threats. The results showed an improved detection time for the primary threat while increased reaction time to the secondary threat which was placed after the primary threat. In another study (*17*), the impacts of dynamic route guidance on work zone safety under different market penetration of CV were explored. Per the interesting results, 40% penetration of CV and below improves safety while above that leads to decreased safety of work zones. However, these benefits are dependent on the information dissemination delay (*18*). Although, positive effects of warning messages have been investigated, the way those warning should be created from BSMs is still under explored.

One approach is trying to link the generation of warning messages to drivers' behavior. In some recent studies, the authors have initiated efforts to extract useful information from BSMs to understand the drivers' behavior. For instance, a measure of driving performance in connected vehicles network has been defined as "driving volatility" (*19*). As such, trip-based driving volatility was introduced (*19*) to account for the variation of driving behaviors under different conditions using objective driving performance evaluation matrix i.e. vehicular jerk. More succinctly, Liu *et al.* (*20*) studied extreme driving behaviors (trip-based volatility) using exhaustive high frequency connected vehicle data, and the analysis demonstrated framework for the generation of warnings/alerts for connected vehicles informing drivers about potential hazards. Also another study (*21*) proposed a way to identify abnormal or extreme behaviors (i.e., hard acceleration and decelerations) from BSMs, and warn drivers through the V2V, V2I, or other connected vehicle applications. In this paper, the authors believe that expanding the concept of driving volatility in connected vehicles environment to specific locations has significant potential in identifying hazardous roadway segments. Such a perspective of location-specific driving behavior in connected vehicle systems has not been identified and analyzed. Therefore, this paper is aimed at developing the new concept of location-based driving volatility (LBV) via using BSMs exchanged between connected vehicles in real-world and linking it to historical crash data with the purpose of identifying hazardous spots proactively. Although the novelty of this study is in using high volume and high velocity connected vehicle data, the significance of works done by other researchers on crash frequency cannot be overlooked, given the emergence of new approaches, e.g., see Lord & Mannering (*22*). Also random parameter and/or varying coefficient models have become popular as opposed to fixed parameter for their capability to address unobserved heterogeneity (*23-25*).

**Research Objective and Contribution**

The objectives of this study are to:

1) Quantify instantaneous driving decisions and its variability in intersection-specific Basic Safety Messages (BSMs).
2) Understand the relationship between intersection-specific volatility with crash frequencies, while controlling for other variables, using rigorous statistical tools.

   The present study contributes by analyzing real-world large-scale connected vehicle data to extract critical driving behavior information embedded in raw BSMs. Such an analysis is important because driving actions and behaviors are believed to be the main cause of traffic crashes, and understanding the relationship between location-based volatility and historical crash data can provide fundamental knowledge regarding proactive safety countermeasures. A unique aspect of this study is that significant efforts have been undertaken to integrate large-scale connected vehicle data (more than 65 million BSMs) with intersection crash and inventory data in order to provide providing a more complete picture of operations and safety performance of intersections. The assembled database allows investigation of correlations between potentially leading indicator of safety (location-based volatility) and historical crash frequencies. By taking the first step towards proactive safety using large-scale connected vehicle data, the current study is original and timely in sense that real-world data has been processed and used to understand the phenomena under discussion.

# METHODOLOGY
## Conceptual Framework
The two-month connected vehicle data from Safety Pilot Model Deployment (SPMD) (https://www.its-rde.net/home) contains rich information (i.e., basic safety messages in 10 Hz) that was exchanged between vehicles/infrastructure through vehicle-to-vehicle (V2V) and vehicle-to-infrastructure (V2I) applications. Such data provide us with an opportunity to scrutinize the mechanisms that lead to unsafe events on roadways. However, the methods of making a good use of such high-volume and high-resolution data need further development. SPMD collects Basic Safety Messages (BSMs) that describe a vehicle's position, motion, its component status, and other relevant travel information (*26*). However, BSMs are not informative to drivers when they need to make decisions based on information received through V2V or V2I applications. Most BSMs describe normal driver behaviors while abnormal and highly fluctuating driver behaviors determine the safety of driving in the short-term.

This study is focused on developing an innovative methodology for estimating location-based volatility for specific intersections and comparing it with their observed crashes. We hypothesized that the nature of extreme instantaneous driving behaviors at intersections can be correlated with their crash history. Such correlations can help us understand instantaneous driving behaviors and how they relate to transportation safety. Location-based volatility (LBV) represents the driving performance of a substantial number of users traveling through a specific location. LBV may play a critical role in highway safety management, as it will highlight locations where many drivers behave differently from other locations. Proactive countermeasures can be considered in such locations. If many drivers make extreme driving behaviors or if driving behaviors are highly fluctuating at certain locations, the reasons of such extreme behaviors may be related to factors such as the road conditions. Such information can be disseminated to connected vehicle drivers through roadside equipment (RSE) which are able to send information to vehicles, and thus drivers may be alerted about potential hazards (e.g. conflicts/intersection sight distance) while traveling through certain intersections.

First, the connected vehicle data consisting of geo codes and longitudinal acceleration were cleaned. In the next step, 116 intersections were identified in Ann Arbor, Michigan (discussed later). Crash data along with other geometric elements (provided in Table 1) were collected. Then, four different coefficients of variation ($C_{V_{AL}}, C_{V_{AH}}, C_{V_{DL}}, C_{V_{DH}}$) are calculated and used as measures of location-based volatility (LBV) for each intersection (150 ft. from the center of each intersection). Given the hypothesis that higher LBV is likely to be positively correlated with historical crash data at intersections, appropriate statistical models are developed to investigate the correlation between LBV (among other traffic exposure factors) and crash frequency. The knowledge generated from the modeling results can identify intersections where drivers, on average, show higher volatility in their instantaneous driving decisions (e.g. longitudinal acceleration), and where such volatilities are found to be correlated with crash frequency. By carefully analyzing high-resolution real-world data transmitted between connected vehicles and applying appropriate statistical methods, we can ultimately generate proactive (rather than the traditionally reactive safety approach) alerts and warnings given to vehicle drivers at intersections. Such proactive warning and alerts can be disseminated through roadside equipment to vehicles approaching specific intersections to warn them regarding the chance or ranking of intersection in terms of crash occurrence. In the next section, the computation of LBV is discussed.

**Location Based Volatility**

Understanding instantaneous driving volatility at specific intersections is one of the most challenging aspects of the current study. To calculate location-based volatility, different instantaneous driving measures can be used such as accelerations, steering angles or position of brakes (*6*). As explicitly discussed in Liu and Khattak (*6*), volatility in trip-based instantaneous driving decisions should be captured by considering both longitudinal and lateral accelerations. Considering longitudinal acceleration as the only measure of driving volatility can mask important information embedded in instantaneous driving data. For instance, at moments longitudinal acceleration can be low and thus considered normal, but the driver could still be volatile due to large magnitudes of lateral accelerations.

To calculate LBV, the authors intended to use longitudinal and lateral acceleration as they are direct outcomes of vehicle maneuvering. However, due to a considerable amount of questionable lateral acceleration data (see Data Accuracy section), only longitudinal acceleration data were used. The longitudinal acceleration data is reasonable and available for all BSMs and has been error checked by estimating accelerations from speed trajectories of the vehicles. Given the data limitation, this study only focuses on capturing location-based volatility by using longitudinal accelerations. There are two reason for this decision: First, excluding lateral acceleration does not seem to be affecting the results drastically since lateral acceleration is more informative in trip based volatility calculation where curvature of the road changes and where the length of the trip allows several lane changes. Second, using the data with removed lateral acceleration reduces the amount of data for several intersections leading to reduction of sample size i.e. number of intersections.

**Calculation of LBV**

The present study uses a standardized measure of dispersion called Coefficient of Variation ($C_V$) (also known as the ratio of relative standard deviation) for quantifying the fluctuations in longitudinal acceleration /decelerations at a specific intersection. Note that different measures such as range, interquartile range, variance or standard deviation can be used for capturing variability in longitudinal accelerations. Although standard deviation and variance are preferable as whole information embedded in the data is used for calculation of variability, both measures are insensitive to magnitude of acceleration values in the data. Thus, we prefer the relative measure of dispersion (Coefficient of Variation), where the dispersion in accelerations or decelerations can be quantified as the proportion of their means. This approach can capture the variability (e.g. standard deviation) in instantaneous driving decisions with respect to the mean accelerations or decelerations undertaken by different drivers at a specific intersection.

To compute volatility for each intersection, two speed bins (see Figure 1a), one from minimum observed speed to the mean and one from the mean to maximum speed were considered. The rationale behind considering speed bins is that the acceleration capability of a vehicle depends on current vehicle speed i.e. at larger speeds the capability to accelerate decrease as compared to acceleration capability at lower speeds. For each bin within an intersection, acceleration and deceleration values are separated, and the means and standard deviations are computed. Finally, $C_V$ as a measure of LBV is obtained by dividing standard deviations of accelerations to the mean, i.e., $\sigma/\mu$. For each intersection, four $C_V$s are reported as shown in Figure 1a. The calculated $C_V$ for a specific intersection provide the relative measure of dispersion of longitudinal accelerations with respect to their means, and thus different intersections can be compared based on their $C_V$s.

**Modeling Approach**

After quantification of volatility for each intersection, we investigate the correlations between location-based volatility (for each intersection), crash data, and other traffic related factors. Appropriate modeling can provide an empirical evidence as of how intersection location-based volatility relates to historical crash data. Given the count nature of crashes, Poisson and/or Poisson-gamma models (Negative Binomial) can be estimated depending on the mean and variance of crash data.

For a Poisson model, the probability of having a specific number of crashes "$n$" at intersection "$i$" can be written as ($27$):

$$P(n_i) = \frac{\exp(-\lambda_i)\lambda_i^n}{n_i!} \tag{1}$$

Where: $P(n_i)$ is probability of crash occurring at intersection "$i$", "$n$" times per specific time-period; and $\lambda_i$ is Poisson parameter for intersection "$i$" which is numerically equivalent to intersection "$i$" expected crash frequency per year $E(n_i)$. The regression can be fitted to crash data by specifying $\lambda_i$ as a function of explanatory variables such as location-based volatility, Annual Average Daily Traffic, and speed limits on major and minor approach. Formally, $\lambda_i$ can be viewed as a log link function of a set of independent variables (27):

$$\ln(\lambda_i) = \beta(X_i) \tag{2}$$

Where $X_i$ is a vector of explanatory variables and $\beta$ is a vector of estimable parameters.

Application of Poisson regression to over-dispersed crash data can result in inappropriate results. If mean and variance of crash data are not equal, corrective measures are applied to Equation 2 by adding an independently distributed error term $\in$. While presence of over-dispersion can be indicated by the mean and variance of crash data (27), formally a Lagrange multiplier can be performed to statistically test the existence of over- dispersion in Poisson model (27). The test statistic is defined as:

$$LM = \left[\frac{\sum_{i=1}^{n}[(y_i - \mu_i)^2 - y_i]}{2\sum_{i=1}^{n}\mu_i{}^2}\right]^2 \tag{3}$$

Where: $y_i$ are actual crash frequency for intersection "$i$", $\mu_i$ is expected crash frequency for intersection "$i$" as predicted by Poisson model, and $n$ are number of observations. The null hypothesis is that Poisson regression is appropriate for the crash data at hand. Under this hypothesis, the LM test statistic should have chi-square distribution with degree of freedom equal one. If the asymptotic chi-square distribution obtained from Equation 3 is less than critical chi-square of 3.84 at 95% level of confidence, Poisson regression should be favored, otherwise Negative Binomial regression can be more appropriate (27).

Finally, it is likely that the associations between key explanatory variables and crash frequency may not be consistent across intersections. The intrinsic unobserved heterogeneity can arise due to several observed and unobserved factors related to intersection crash frequency, which may not be available in the data at hand. This is referred to omitted variable bias in safety literature (27). Furthermore, if key variables are omitted from analysis and too few variables are included in the model, it is likely that location-based volatility (explanatory factor) can capture those effects and may not be the true association between location-based volatility and crash frequency. One way to address this issue is to allow parameter estimates to vary across observations (27). As such, random parameters can be included in the estimation framework as:

$$\beta_i = \beta + \varphi_i \tag{4}$$

Where $\varphi_i$ is randomly distributed term with any pre-specified distribution such as normal distribution with mean zero and variance $\sigma^2$. With Equation 4, the Poisson parameter in Equation 2 becomes:

$$\lambda_i|\varphi_i = e^{BX} \tag{5}$$

And, the Poisson parameter in Equation 2 in Poisson-Gamma model becomes:

$$\lambda_i|\varphi_i = e^{BX+\epsilon} \tag{6}$$

Finally, the following likelihood function for random-parameter model can be maximized through maximum simulated likelihood technique (23):

$$LL = \sum_i ln \int_{\varphi_i}^{i} g(\varphi_i)P(n_i|\varphi_i)d\varphi_i \tag{7}$$

Where: g(.) is the probability density function of randomly distributed term with pre-specified distribution such as normal distribution with mean zero and variance $\sigma^2$. More details on random parameter models can be found in (*23*).

## DATA

The data used in this study (retrieved from https://www.its-rde.net/home) are BSMs from vehicles participating the SPMD in Ann Arbor, Michigan. SPMD is a comprehensive data collection effort, under real-world conditions, at Ann Arbor test site with multimodal traffic hosting approximately 3,000 connected vehicles equipped with V2V and V2I communication devices. BSMs are frequently transmitted messages (usually at 10Hz) that is meant to increase vehicle's situational awareness. At its core, the dataset contains vehicle's instantaneous driving statuses of vehicle's position (latitude, longitude, and elevation) and motion (heading, speed, accelerations).

To examine correlations, location-based volatility (LBV) data for each intersection (as explained earlier) are linked with historical crash data, annual average daily traffic (AADT) data for major and minor approaches, speed limits on major and minor approaches, and number of approaches at each intersection. Such data are publicly available at the website of the Metropolitan Planning Organization: http://semcog.org/Data-and-Maps. Out of all intersections in the Ann Arbor area, 116 intersections are identified for which connected vehicle data are available, i.e. connected vehicles pass through such intersections and generating enough data for calculation of LBV. Finally, five years of crashes (2011-2015) along with geometric factors and flows were extracted and linked to LBV for each intersection. Note that the data are not available in spreadsheet format, and thus significant efforts went into carefully extracting data manually and linking it to LBV for 116 intersections.

### Data Accuracy

Based on the distributions of key variables provided in Table 1, the data seems to be of reasonable quality. To assure the accuracy of intersection data, after initial collection, another person checked 10% of intersection data randomly and no discrepancies were observed. Also, the descriptive statistics of intersection data in Table 1 provide reasonable difference between signalized and un-signalized intersections. The major inaccuracy of data is from the lateral acceleration as it is shown in Figure 1b. Since 27,240,788 data points (42% of the data) had the maximum allowable value that can be recorded in DSRC devices (2g), lateral acceleration data was not used in the analysis.

**(FIGURE 1 HERE)**

### RESULTS

### Descriptive Statistics

Table 1 presents the descriptive statistics of key variables used in modeling. The mean, standard deviation, minimum and maximum values are given for each variable which can help conceptualizing the distributions. Descriptive statistics are given for all the intersections (N=116) as well as for signalized intersections (N=53) and un-signalized intersections (N=63) separately. For all intersections, signalized, and un-signalized intersections, the mean five-year crash frequency is 7.56, 12.94, and 3.04. As expected, signalized intersections have significantly higher crash frequency (on average) than un-signalized intersections. This finding is in agreement with Abdel Aty and Keller (*28*) who found approximately 9.6 crashes per year at signalized intersections as opposed to only 2 crashes per year on un-signalized intersections (*28*). There can be several factors which may contribute to occurrence of crashes at signalized intersections such as conflicting movements as well as different intersection-specific design variables (*28*). This said, investigating instantaneous driving actions at such locations, and higher volatility (if any) may help us design appropriate proactive strategies from preventing an "accident waiting to happen" (*29*).

Regarding location-based volatility, all $C_V$ statistics suggest that signalized intersections on average have higher variability in longitudinal accelerations/decelerations compared with unsignalized intersections, and thus can be more volatile (this is the case for all $C_V$'s except $C_{V_{AH}}$). One reason for higher $C_{V_{AH}}$ (volatility of acceleration above mean speed) of un-signalized intersections as compared to signalized intersections can be due to uninterrupted traffic of un-signalized intersections.

In order to avoid omitted variable bias in modeling (*30*), data on other variables such as five-year average AADT (major and minor approach), speed limits (major and minor approaches), and number of approaches were collected. Regarding the number of approaches, 40% of all intersections, 62.2% of signalized intersections, and 22% of un-signalized intersections are four-legged intersections (Table 1). In terms of exposure on major and minor roads, signalized intersections have higher (on average) AADT than un-signalized intersections (22,747 vs. 19,171 for major roads and 9,994 vs. 8,893 for minor roads). Regarding number of lanes, number of through and left turns for signalized intersection are considerably higher as compared to un-signalized intersections.

<p style="text-align:center">**(TABLE 1 HERE)**</p>

## Modeling Results

For examining the correlations between crash frequency and location-based volatility (as measured by $C_V$s), count data models are estimated given the count nature of crash frequency. Separate count data regression models are estimated for all intersections, signalized intersections and un-signalized intersections. Specifically, fixed-parameter Poisson regressions are estimated for total crash frequency as a function of location based volatility, major and minor road AADT, major and minor road speed limits, and total number of through lanes. It should be noted that the descriptive statistics for crash frequencies in Table 1 apparently reveal the existence of over-dispersion in the data where Negative Binomial model should be preferred over Poisson model (*31*).Thus, statistical tests are conducted to confirm the existence of over-dispersion (*27*). As explained in methodology section, Lagrange Multiplier tests were conducted for all three Poisson models. By using Equation 3, the Lagrange Multiplier (LM) values were 0.05, 0.031, and 0.15 for all intersections, signalized intersections, and un-signalized intersections respectively. The LM values are much smaller than critical Chi-square value of 3.84 for one degree of freedom at 95% confidence level. Thus, the null hypothesis that Poisson regressions are more appropriate is failed to reject, and it would be more appropriate to use Poisson regressions (*31*).

Due to the likely presence of unobserved heterogeneity in crash data (*23*) which may arise due to several unobserved factors, random-parameter Poisson models are also estimated. Fixed parameter models are estimated with standard maximum likelihood whereas random parameter models are estimated through simulated maximum likelihood with 200 Halton draws used for random-held parameters (*23*). Regarding functional form of random-parameters, log-normal, Weibull, uniform, and triangular distributions are tested with normally distributed random parameters giving the best fit and shown in this study.

The results obtained from fixed and random parameter Poisson model are presented in Table 2. Marginal effects are also provided for the random parameter models that translate unit change in crash frequency with unit change in explanatory variable. Compared to fixed-parameter models, random-parameter models resulted in better fit as of improved log-likelihood at convergence and McFadden's $\rho^2$ (Table 2) (*31*). While this study does not focus on methodological approaches for modeling intersection crash data, the predicted vs. actual values of crashes (Figure 2) are plotted and reveal statistical superiority of random parameter models in fitting the data at hand.

## Discussion

Coming to the fixed-parameter estimation results for all intersections (Table 2), the results provide evidence that $C_{V_{AL}}$, $C_{V_{DL}}$, and $C_{V_{DH}}$ are positively associated (statistically significant at 95% confidence level) with crash frequency. However, $C_{V_{AH}}$ is negatively associated with crash frequency (at 90% confidence interval). It can be concluded, overall, volatility of deceleration regardless of speed range is positively associated with crash frequency. However, when it comes to acceleration, volatility at lower speed is more a significant factor as compared to volatility at higher speeds.

At signalized intersections, the association between $C_{V_{AL}}$, $C_{V_{AH}}$ and $C_{V_{DH}}$ and crash frequency is also positive and statistically significant. $C_{V_{DL}}$ for signalized intersection; however, it is negatively correlated with crash frequency.

Referring to marginal effects for random parameter model in Table 2, on average one-percent increase in $C_{V_{DH}}$ is associated with 0.11 increase in crash frequency for all intersections and 0.089 increase in crash frequency for signalized intersections. These findings have implications for proactive intersection-related safety strategies. In addition, it is interesting to note the significantly higher marginal effect of acceleration $C_V$s for signalized intersections, implying that higher variability in acceleration at signalized intersections may potentially result in more crashes. Given that signalized intersections are typically observed to have more crashes (*28*), proactive intersection-customized strategies can be designed. For instance, proactive warnings and alerts can be generated about potential hazards at specific intersections and transmitted to drivers via connected vehicle technologies such as road-side equipment. This can in turn increase drivers' situational and safety awareness, and help drivers in undertaking safer driving behaviors.

Regarding un-signalized intersections, as shown in Table 2, $C_{V_{AL}}$ and $C_{V_{DL}}$ are statistically significant. We found negative association between $C_{V_{AL}}$ and crash frequency. This finding is seemingly counter intuitive and needs further investigation. Possibly, for un-signalized intersection, due to their uninterrupted traffic in major approach (78% of them are T-intersections), separation of 3-leg and 4-leg intersection might shed more clarification in future studies. However, the finding that $C_{V_{DL}}$ (Coefficient of variation of deceleration below mean speed of intersection) is positively associated with crash frequency is intuitive i.e. larger the volatility/variation in decelerations at low speeds, the more crash frequency at a particular intersection.

The estimation results quantify associations between major and minor road AADT and crash frequency. Referring to marginal effects from the random-parameter model, one-log unit increase in major road AADT is associated with 2.69, 6.57, and 1.82-unit increase in crash frequency for all intersections, signalized intersections, and un-signalized intersections, respectively. Minor road AADT is statistically significant in the random-parameter model for signalized intersections, but the relationships are not statistically significant for un-signalized intersections (Table 2). Speed limit on major roads is negatively associated with crash frequency for all intersections. These findings are consistent with past studies on this topic (*1; 32*). Notably, the total number of through lanes is positively associated with crash frequency. From Table 2, it can be observed that one added through lane is correlated with 0.547 more crashes.

Figure 3 illustrates how the study results can assist in proactive intersection safety management. The black, green and red circles in the figure are scaled crashes, volatility of acceleration, and volatility of deceleration at lower speeds, respectively. The intersection in the center is a known hotspot because it has more crashes and proportionately high levels of volatility. However, two other intersections shown in dashed ellipses have relatively low crashes but high volatility levels ($C_{V_{AL}}$, $C_{V_{DL}}$). In such locations (hotspots), although crash frequencies are low, drivers show proportionately more volatile driving behavior. In other words, at such locations crashes may be waiting to happen. Proactive countermeasures can be taken in those locations depending on the real cause of driving volatility, e.g., by studying speed limits, signal timing, geometric design, dilemma zone, and lines of sight.

**(FIGURE 2 HERE)**

**(FIGURE 3 HERE)**

## LIMITATIONS
The study captures variability in longitudinal acceleration/deceleration as a measure of intersection-specific volatility, which only partially capture the true volatility exhibited by drivers. As explained in the methodology section, due to data limitations, the study could not incorporate lateral acceleration/deceleration in estimation of intersection-specific volatility. While the results from this study provide evidence between crash frequency and intersection-specific volatility, more robust measures such as vehicular jerk and combination of longitudinal and lateral accelerations can be used in future studies for quantifying volatility at specific intersections. Also, the results and conclusions of this study are dependent on the sample-size. Another limitation is that one month data were used to explain 5-year average crash. While the current sample size may not be enough to draw robust conclusions, the authors have used all available data for 116 intersections.

## CONCLUSIONS
This study contributes by developing and demonstrating a proactive intersection safety methodology using real-world large-scale connected vehicle data. The study quantifies volatility in instantaneous driving decisions using intersection-specific Basic Safety Messages (BSMs) and its relationship with observed crash frequencies, while controlling for other variables. Such a method can complement the state-of-the-art in evaluating intersection safety, which is largely reactive, based on observed and expected crash frequencies. The emerging data from Connected and Automated (CAVs) are increasingly becoming available, which can help us understand the detailed nature of instantaneous driving behaviors prior to the occurrence of unsafe outcomes such as crashes/incidents. This study proposes the concept of location-based volatility that captures the extent of variations in instantaneous driving decisions.

A unique database that provides a more complete picture of operations and safety performance was created by combining more than 65 million Basic Safety Messages transmitted between connected vehicles and roadside units at 116 intersections in Ann Arbor, Michigan, with crash and inventory data. The geo-coded raw BSMs were allocated to each intersection and the connected vehicles trajectories extracted from raw BSMs were plotted, revealing reasonable data precision and coverage. A simple and standardized measure of dispersion called Coefficient of Variation ($C_V$) (also known as the ratio of relative standard deviation) was used to quantify the fluctuations in longitudinal acceleration and/or decelerations at specific intersections. Five-year crash frequencies, AADT, speed limits, and number of approaches for all intersections are extracted and linked with location-based volatilities. Significant efforts went into data processing, collection, and linkage.

Rigorous fixed and random parameter Poisson regression models are estimated that allow consideration of unobserved heterogeneity in crash data. The modeling results reveal that most of computed $C_V$s (as measures of volatilities) are positively associated with crash frequency. The study has implications for proactive intersection safety management. Importantly, the magnitude of association between location-based volatility and crash frequency is significantly higher for signalized intersections, implying that higher variability in instantaneous driving decisions at signalized intersections may potentially result in more crashes. This finding is important in the sense that if many drivers behave in a volatile manner at a specific intersection (exhibit higher variability in longitudinal accelerations), then such intersections can be identified before accidents happen. Of course, the reasons for volatile behaviors may be related to intersection and environmental conditions, vehicles' and drivers' conditions. Given that signalized intersections are typically observed to have more crashes (*28*), intersection-customized strategies can be designed to improve safety. Proactive warnings and alerts can be generated about potential hazards at specific intersections and transmitted to drivers via connected vehicle technologies such as road-side equipment; these can in turn increase drivers' situational and safety awareness, and help them pursue safer driving at dangerous intersections.

## ACKNOWLEDGEMENT

This paper is based upon work supported by the US National Science Foundation under grant No. 1538139. Additional support was provided by the US Department of Transportation through the Collaborative Sciences Center for Road Safety, a consortium led by The University of North Carolina at Chapel Hill in partnership with The University of Tennessee. Any opinions, findings, and conclusions or recommendations expressed in this paper are those of the authors and do not necessarily reflect the views of the sponsors.

LIST OF TABLES

LIST OF FIGURES

**TABLE 1 Description of Key Variables and Descriptive Statistics**

| Variables | All Intersections (N = 116) | | | Signalized (N = 53) | | | Un-signalized (N=63) | | |
|---|---|---|---|---|---|---|---|---|---|
| | Mean | SD | Min/Max | Mean | SD | Min/Max | Mean | SD | Min/Max |
| Average crashes (5 years) | 7.56 | 7.64 | 0/44 | 12.94 | 8.03 | 1/44 | 3.04 | 2.95 | 0/14 |
| Average rear-end crashes (5 years) | 4.28 | 4.56 | 0/24 | 7.07 | 5.24 | 1/24 | 1.93 | 1.79 | 0/9 |
| $C_{V_{AL}}$ (In percent) | 143.71 | 56.03 | 69/239 | 182.44 | 57.58 | 83/329 | 111.13 | 26.12 | 69/191 |
| $C_{V_{AH}}$ (In percent) | 84.9 | 13.76 | 56/121 | 77.93 | 12.7 | 59/113 | 90.77 | 11.8 | 57/121 |
| $C_{V_{DL}}$ (In percent) | 137.51 | 43 | 71/287 | 168.67 | 41.15 | 87/287 | 111.29 | 21.94 | 71/181 |
| $C_{V_{DH}}$ (In percent) | 96.29 | 12.9 | 57/155 | 99.44 | 14.86 | 76/155 | 93.64 | 10.39 | 57/115 |
| AADT major road | 20805 | 8326 | 3100/45400 | 22747 | 8209 | 3600/45400 | 19171 | 8131 | 3100/38900 |
| AADT minor road | 9396 | 4138 | 1100/27400 | 9994 | 5706 | 3100/27400 | 8893 | 1972 | 1100/13400 |
| Ln (AADT major road) | 9.84 | 0.49 | 8.03/10.72 | 9.96 | 0.39 | 8.18/10.72 | 9.74 | 0.54 | 8.03/10.56 |
| Ln (AADT minor road) | 9.05 | 0.47 | 7/10.21 | 9.07 | 0.52 | 8.03/10.21 | 9.03 | 0.42 | 7/9.50 |
| Speed limit major | 35.34 | 7.24 | 25/45 | 35.94 | 7.34 | 25/45 | 34.84 | 7.18 | 25/45 |
| Speed limit minor | 30.47 | 3.95 | 25/45 | 30.84 | 5.16 | 25/45 | 30.15 | 2.53 | 25/40 |
| 4-legged intersection | 0.4 | 0.49 | 0/1 | 0.622 | 0.489 | 0/1 | 0.22 | 0.41 | 0/1 |
| Total through lanes | 4.45 | 1.28 | 2/8 | 5.13 | 1.35 | 2/8 | 3.38 | 0.9 | 2/6 |
| Total left turn lanes | 1.53 | 1.32 | 0/6 | 2.26 | 1.4 | 0/6 | 0.92 | 0.88 | 0/3 |
| Total right turn lanes | 0.93 | 0.78 | 0/4 | 1.11 | 1.01 | 0/4 | 0.79 | 0.48 | 0/2 |

Notes: $C_{V_{AL}}$: Coefficient of variation of acceleration below mean speed of intersection; $C_{V_{AH}}$: Coefficient of variation of acceleration above mean speed of intersection; $C_{V_{DL}}$: Coefficient of variation of deceleration below mean speed of intersection; $C_{V_{DH}}$: Coefficient of variation of deceleration above mean speed of intersection; AADT: Annual Average Daily Traffic; SD is standard deviation; Min is minimum value; Max is maximum value.

**TABLE 2 Modeling Results of Fixed- and Random-Parameter Poisson Regressions**

| Variables | Signalized and Un-signalized | | | | | Signalized Intersections | | | | | Un-signalized Intersections | | | | |
|---|---|---|---|---|---|---|---|---|---|---|---|---|---|---|---|
| | Fixed Par. | | Random Par. | | | Fixed Par. | | Random Par. | | | Fixed Par. | | Random Par. | | |
| | $\beta$ | t-stat | $\beta$ | t-stat | ME | $\beta$ | t-stat | $\beta$ | t-stat | ME | $\beta$ | t-stat | $\beta$ | t-stat | ME |
| Constant | -7.752 | -6.6 | -7.786 | -7.237 | --- | -7.21 | -4.975 | -7.35 | -6.958 | --- | -10 | -3.574 | -9.61 | -3.235 | --- |
| *Standard deviation** | --- | --- | --- | --- | --- | --- | --- | --- | --- | --- | --- | --- | 0.488 | 6.155 | --- |
| $C_{V_{AL}}$ | 0.006 | 4.152 | 0.004 | 2.902 | 0.025 | 0.009 | 3.434 | 0.01 | 5.346 | 0.125 | -0.014 | -2.831 | -0.016 | -2.911 | -0.035 |
| *Standard deviation* | --- | --- | --- | --- | --- | --- | --- | 0.0002 | 1.991 | --- | --- | --- | --- | --- | --- |
| $C_{V_{AH}}$ | -0.003 | -0.776 | -0.007 | -1.983 | -0.038 | 0.009 | 1.453 | 0.01 | 1.959 | 0.118 | 0.005 | 0.683 | 0.004 | 1.28 | 0.01 |
| *Standard deviation* | --- | --- | 0.005 | 11.856 | --- | --- | --- | --- | --- | --- | --- | --- | --- | --- | --- |
| $C_{V_{DL}}$ | 0.002 | 1.243 | 0.005 | 2.827 | 0.027 | -0.003 | -1.541 | -0.004 | -2.222 | -0.057 | 0.015 | 2.698 | 0.0153 | 3.186 | 0.036 |
| *Standard deviation* | --- | --- | --- | --- | --- | --- | --- | 0.0009 | 4.363 | --- | --- | --- | --- | --- | --- |
| $C_{V_{DH}}$ | 0.02 | 6.449 | 0.021 | 6.33 | 0.11 | 0.008 | 1.872 | 0.007 | 1.981 | 0.089 | -0.0007 | -0.09 | 0.0001 | 0.05 | 0.0004 |
| *Standard deviation* | --- | --- | 0.0007 | 2.182 | --- | --- | --- | --- | --- | --- | --- | --- | --- | --- | --- |
| Ln (Major Road AADT) | 0.547 | 4.899 | 0.527 | 5.322 | 2.694 | 0.55 | 3.716 | 0.565 | 5.561 | 6.575 | 0.866 | 4.801 | 0.757 | 4.106 | 1.823 |
| *Standard deviation* | --- | --- | 0.011 | 3.376 | --- | --- | --- | --- | --- | --- | --- | --- | 0.488 | 6.155 | --- |
| Ln (Minor Road AADT) | 0.123 | 1.656 | 0.15 | 1.97 | 0.767 | 0.191 | 2.083 | 0.207 | 2.03 | 2.413 | 0.231 | 1.004 | 0.292 | 1.25 | 0.704 |
| *Standard deviation* | --- | --- | 0.006 | 2.152 | --- | --- | --- | --- | --- | --- | --- | --- | --- | --- | --- |
| Speed limit major road | -0.009 | -1.736 | -0.014 | -2.497 | -0.073 | 0.004 | 0.576 | 0.008 | 1.227 | 0.097 | --- | --- | --- | --- | --- |
| Speed limit minor road | --- | --- | --- | --- | --- | -0.016 | -1.444 | -0.023 | -1.62 | -0.271 | --- | --- | --- | --- | --- |
| Total through lanes | 0.61 | 1.733 | 0.107 | 3.223 | 0.547 | --- | --- | --- | --- | --- | --- | --- | --- | --- | --- |
| **Summary Statistics** | | | | | | | | | | | | | | | |
| Log-lik. at Zero $L(0)$ | -578.31 | | -578.31 | | | -226.73 | | -226.73 | | | -158.18 | | -158.18 | | |
| Log-lik. at Convergence $L(\beta)$ | -336.72 | | -305.02 | | | -159.43 | | -154.91 | | | -138.26 | | -130.44 | | |
| McFadden $\rho^2$ | 0.417 | | 0.831 | | | 0.31 | | 0.893 | | | 0.125 | | 0.59 | | |
| Sample Size (N) | 116 | | | | | 53 | | | | | 63 | | | | |

Notes: ME: Average Marginal Effects from Random Parameter Model. $C_{V_{AL}}$: Coefficient of variation of acceleration below mean speed of intersection; $C_{V_{AH}}$: Coefficient of variation of acceleration above mean speed of intersection; $C_{V_{DL}}$: Coefficient of variation of deceleration below mean speed of intersection; $C_{V_{DH}}$: Coefficient of variation of deceleration above mean speed of intersection; AADT: Annual Average Daily Traffic; *Standard deviation of normally distributed random parameters.

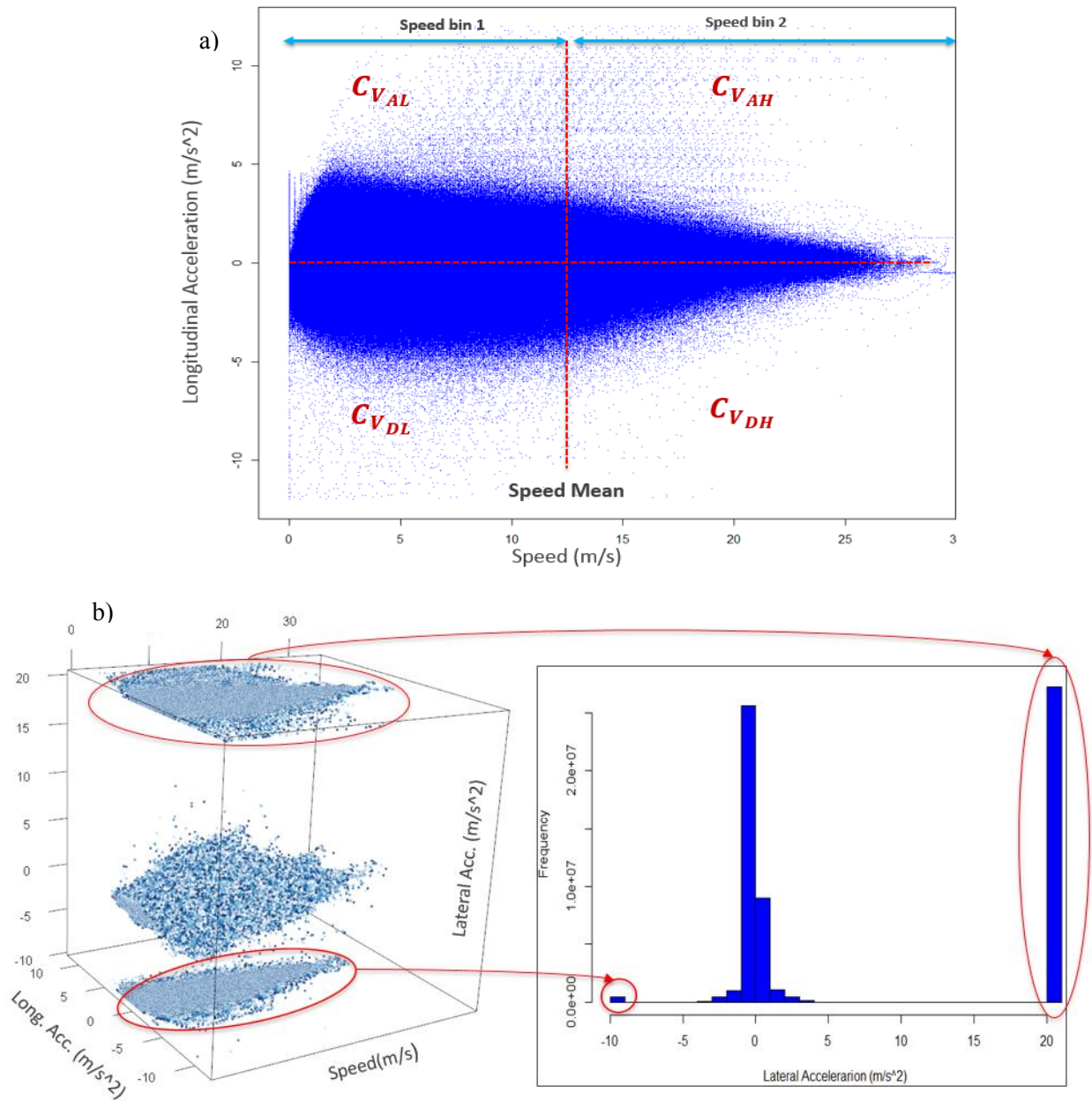

**FIGURE 1: a) Four quadrants used to calculate coefficients of variation (=$\sigma/\mu$) for each intersection, b) Plot of used data (left)/ Histogram of lateral acceleration (right)**

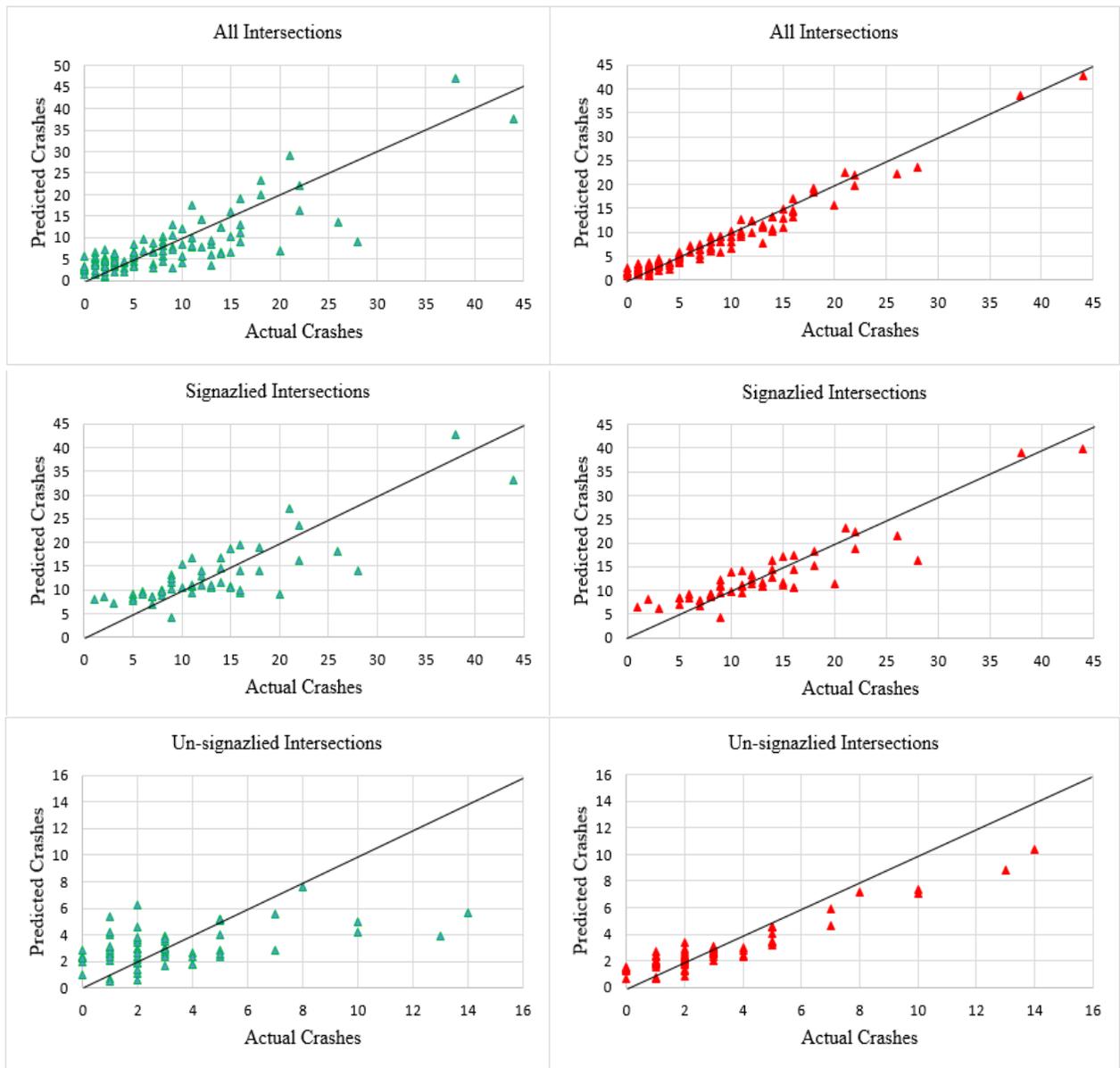

**FIGURE 2: Mean-expected over actual number of crashes for fixed and random-parameter Poisson models (Green: fixed parameter models; Red: random parameter models)**

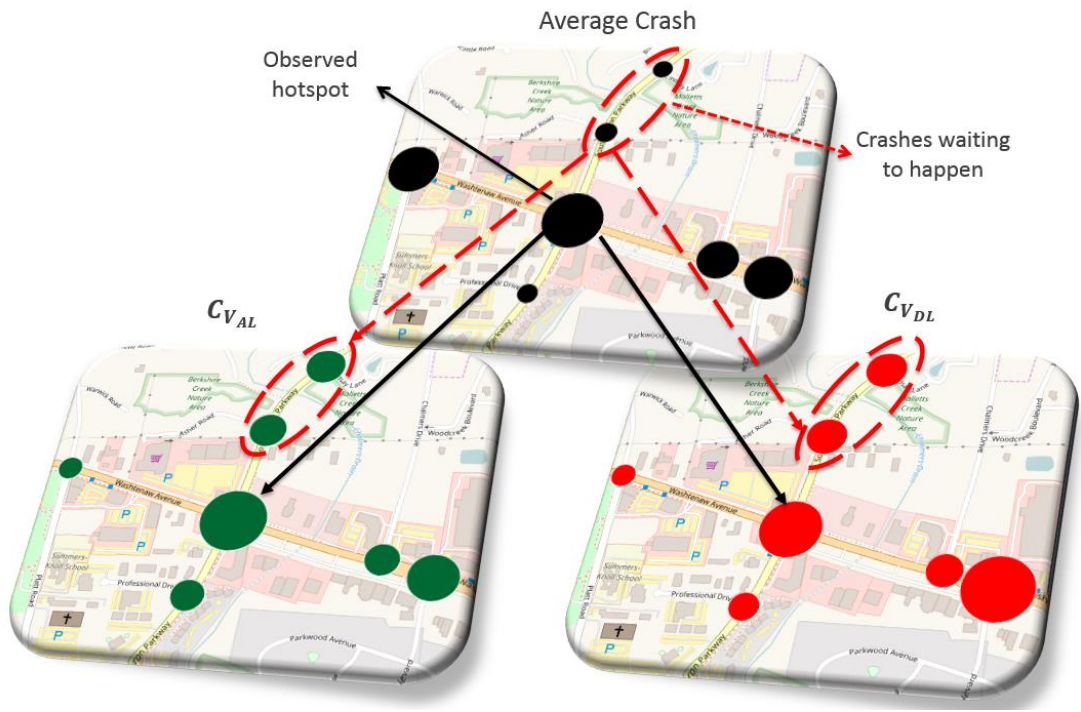

**FIGURE 3: Known hotspots and spots where crashes are waiting to happen.**